\documentclass[epjST]{svjour}
\usepackage{graphicx}
\begin{document}
\title{Non-ergodic states induced by impurity levels in quantum spin chains}
\author{A. O. Garc\'{i}a Rodr\'{i}guez  \and G. G. Cabrera \thanks{\email{cabrera@ifi.unicamp.br}}}
\institute{Instituto de F\'{i}sica \emph{Gleb Wataghin}, Universidade Estadual de Campinas, UNICAMP, 13083-859, Campinas, SP, Brazil}
\abstract{The semi-infinite \emph{XY} spin chain with an impurity at the boundary has been chosen as a prototype of interacting many-body systems to test for non-ergodic behavior. The model is exactly solvable in analytic way in the thermodynamic limit, where energy eigenstates and the spectrum are obtained in closed form.
In addition of a continuous band, localized states may split off from the continuum, for some values of the impurity parameters. In the next step, after the preparation of an arbitrary non-equilibrium state, we observe the time evolution of the site magnetization. Relaxation properties are described by the long-time behavior, which is estimated using the stationary phase method. Absence of localized states defines an ergodic region in parameter space, where the system relaxes to a homogeneous magnetization. Out of this region, impurity levels split from the band, and localization phenomena may lead to non-ergodicity.
}
\maketitle
\section{Introduction}
\label{intro}
    The fundamental assumption of quantum ergodicity asserts that arbitrary initial states should relax to equilibrium at asymptotically long times \cite{vonneumann,comment}. Recently, it has been argued that \emph{ergodicity} should be replaced by \emph{typicality}, in order to justify the approach to the thermodynamic equilibrium. The concept was developed following the old proposal by von Neumann, to base the foundations of Statistical Mechanics in terms of the quantum dynamics of isolated systems. Under some general settings, a situation that has been named as \emph{typical}, it has been shown that single pure initial states do approach thermodynamic equilibrium without invoking ergodicity \cite{bochieri,typical,reimann,tasaki}. Exceptional states do exist, but are overwhelmed by the great majority of typical states that finally warrants the relaxation process. Macroscopic quantities are obtained as quantum averages over a single typical state, and their values are `very close' to those obtained through the conventional ensemble theory.

As pointed out by Tasaki in \cite{tasaki}, the reversible unitary time evolution in isolated quantum systems can describe the process of thermalization, which may appear as being irreversible in the thermodynamic limit. The above limit is a crucial point in the deduction, since quantum recurrences are avoided when the system has a continuous energy spectrum, at asymptotically very large volume \cite{recurrence}. This understanding is largely satisfactory, it applies to individual systems, solves the longstanding paradox of obtaining irreversibility from reversible dynamics, and ergodicity is not necessary for obtaining microcanonical distributions. At the same time, one can predict when thermalization is not expected. It is known that certain isolated systems fail to thermalize, or relax to states other than the equilibrium state \cite{localization}.

At any rate, states that do not relax to equilibrium at very long times, are called non-ergodic. To better understand the above phenomena, one can test non-equilibrium properties in models that are exactly solvable, and eventually find evidence of non-ergodic behavior in a controllable way. Quantum spin chains are non trivial interacting systems that satisfy the above requirements. In this contribution, we discuss examples of non-ergodicity, studying the time evolution of arbitrary initial states in a quantum semi-infinite \emph{XY} spin chain, with an impurity at the boundary. The impurity is characterized by two parameters, its magnetic moment and the coupling with the chain, which may be different from those in the bulk of the system. In the thermodynamic limit, the spectrum has a continuous branch, that we call `the band', and depending on values of the impurity parameters, up to two localized states may be split off from the band. The presence of localized levels in the thermodynamic limit, is a key point for non thermalization.

The problem has been approached before in some different contexts, either using approximations or studying the time-dependent autocorrelation function in finite systems \cite{tjon,stolze}. Advantages of our formulation include: i) solving the problem in exact analytic way using the diagonalization method of Ref.~\cite{tsukernik} in the thermodynamic limit, thus our solution is free of finite-size effects; ii) explore the full parameter space to give a complete account of the dynamical properties; and iii) study of the relaxation process directly through the calculation of the time-dependent average $\left\langle S_{n\vphantom1}^{\mathstrut z}\right\rangle_{\!\!t}$ that yields the evolution of the magnetization, independently of how far the initial state is from the equilibrium state.
This time evolution displays strong oscillations, but most of them are quickly damped by destructive interference. Neighborhoods of stationary points in the spectrum lead to constructive interference, with slow relaxation properties, as in the examples shown in Ref.~\cite{berimPRB}. In the absence of impurity levels, the above relaxation leads to a final state where the magnetization is stationary and homogeneous, and which we identified as being the equilibrium state. This result is what one expects as \emph{typical} or ergodic behavior. Non-ergodic time evolution is observed, when localized impurity levels are split off from the band.

The description of this non-ergodicity is the central matter of our paper, which is organized as follows: in the next Section, we discuss the model used in our calculation, along with the method of solution; after this, we schematically describe the time evolution for a particular (but typical) inhomogeneous initial state; results for the long-time behavior of the magnetization are displayed in the last Section, where we state the final conclusions.
\section{Model calculation}
\label{sec:1}
\begin{figure}[ht!]
\centering
\includegraphics {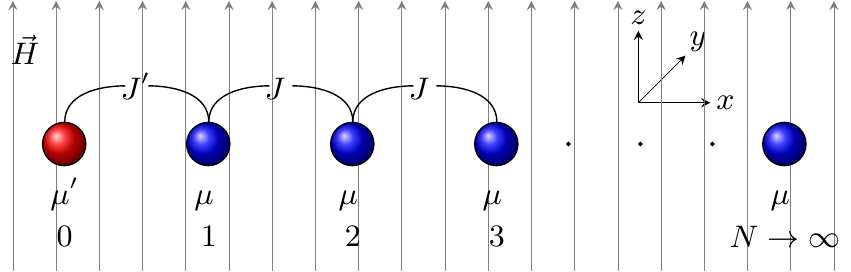}
\caption{Semi-infinite quantum spin chain with an impurity `atom' at site $n=0$. A uniform external magnetic field $H$ is applied along the \(z\)-axis.}
\label{chain}
\end{figure}

We consider a semi-infinite quantum spin chain with an impurity at one end of the chain, as depicted in Fig.~\ref{chain}.

\begin{figure}[ht!]
\includegraphics [width=\linewidth]{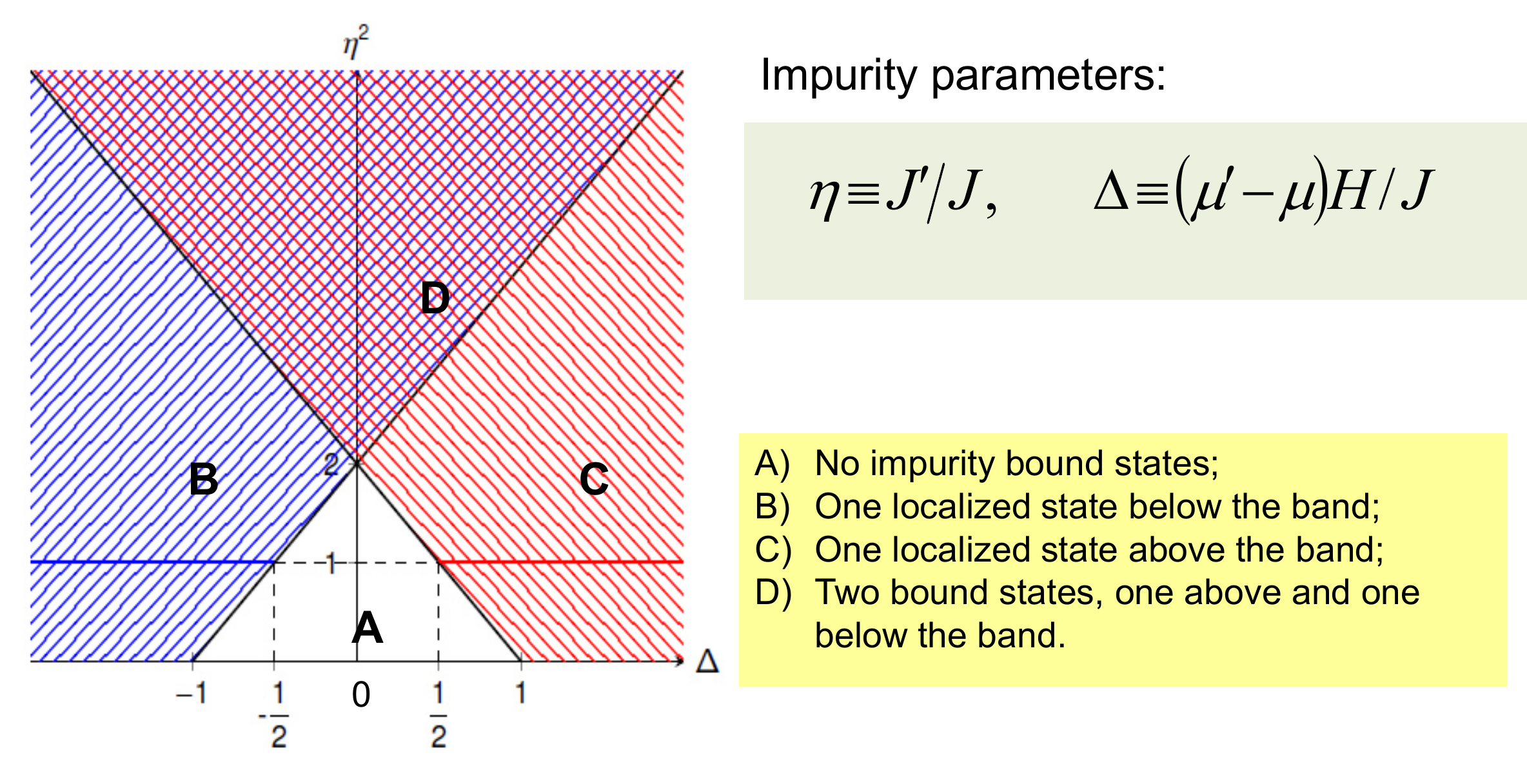}
\caption{Characteristic regions for the presence of localized levels in the spectrum. Dashed and continuous lines are drawn as a guide for the eye. Locus of the pure chain is $\Delta=0$ and $\eta^2=1$. More explanations in the main text.}
\label{levels}
\end{figure}

The figure is self-explanatory, $\mu$ is the magnetic moment in the bulk of the chain, where spins are coupled with exchange $J$. The corresponding quantities for the impurity are $(\mu',J')$. For convenience, to follow the solution of Ref.~\cite{tsukernik}, we define the parameters:
\begin{equation}
\eta=J'/J\qquad\Delta=\left(\mu'-\mu\right)H/J~.
\label{parameters}
\end{equation}
We model our spin chain using the \emph{XY}-Hamiltonian for spin $s=1/2$ with nearest neighbor interactions, in the presence of a uniform magnetic field $H$:
\begin{equation}
\mathcal H = -J'\left(S_0^{\mathstrut x} S_1^{\mathstrut x}\!+\!S_0^{\mathstrut y} S_1^{\mathstrut y}\right)\!-\!J\!\!\sum_{n=1}^{N-1}\!\!\left(S_{n\vphantom1}^{\mathstrut x} S_{n+1}^{\mathstrut x}\!+\!S_{n\vphantom1}^{\mathstrut y} S_{n+1}^{\mathstrut y}\right)-\mu' H S_0^{\mathstrut z}-\mu H\sum_{n=1}^N S_{n\vphantom1}^{\mathstrut z},
\label{ham}
\end{equation}
and we consider the case of a semi-infinite spin chain (\(N\to\infty\)) for which the hamiltonian (\ref{ham}) can be exactly diagonalized~\cite{tsukernik}. The infinite number of degrees of freedom avoids revivals of the initial state, which may be present in finite size systems~\cite{recurrence}. For completeness, we briefly summarize the procedure of Ref. \cite{tsukernik}. We firstly use the Jordan-Wigner transformation, to express spin-1/2 operators in terms of fermion annihilation and creation operators \(c_n^{\mathstrut}\) and \(c_n^\dag\):
\begin{equation}
S_0^+=c_0^\dag,\qquad S_n^+=c_n^\dag\prod_{l=0}^{n-1}\left(1-2c_{l}^\dag c_{l}\right),\quad n\geq 1,
\label{jw}
\end{equation}
where $S_n^+$ are the spin ladder operators. The resulting fermion Hamiltonian is written as:
\begin{equation}
\begin{array}{r}
\mathcal{H}\mathcal{=-} J/2 \sum_{n=1}\ \left( c_{n}^{\dag }\
c_{n+1}+c_{n+1}^{\dag }c_{n}\right) -J^{\prime}/2
\left(c_{0}^{\dag }\ c_{1}+c_{1}^{\dag }\ c_{0}\right) - \\
-\mu H\sum_{n=1}\ \left( \frac{1}{2}-c_{n}^{\dag }\ c_{n}\right) -\mu
^{\prime }H\left( \frac{1}{2}-c_{0}^{\dag }\ c_{0}\right) \ ,%
\end{array}
\label{fermionH}
\end{equation}
which is quadratic in the Jordan-Wigner operators. Hamiltonian (\ref{fermionH}) can be diagonalized by a unitary transformation that preserves the anti-commuting properties:
\begin{equation}
c_n^{\mathstrut}=\sum_{\lambda=1} u_{n\vphantom1}^{(\lambda)} b_\lambda^{\mathstrut},\quad n\geq 0,
\label{transfunit}
\end{equation}
The coefficients $u_{n\vphantom1}^{(\lambda)}$ (wave functions) in (\ref{transfunit}) are derived from the Heisenberg equations of motion for the operators $c_n$, assuming that the Hamiltonian is diagonal in the quasi-particle operators $b_\lambda^{\mathstrut}$:
\begin{equation}
\mathcal H=\sum_{\lambda=1} \epsilon_\lambda~ b_\lambda^\dag\ b_\lambda ,
\label{diagonal}
\end{equation}
with $\lambda$ being the `good quantum number' characterizing the energy spectrum. When the diagonal form (\ref{diagonal}) is attained, the equations of motion for the $b_\lambda^{\mathstrut}$ operators are simply given by:
\begin{equation}
i\hbar\dot{b_\lambda}=[b_\lambda,~\mathcal H]=\epsilon_\lambda~ b_\lambda~,
\label{dotb}
\end{equation}
where $[..,..]$ is the commutator. Using relations (\ref{transfunit}) and (\ref{dotb}), the equation of motion for the $c_n $ operators takes the form:
\begin{equation}
\left[ c_{n},\mathcal{H}\right] =\sum_{\lambda }\epsilon _{\lambda
}u_{n}^{(\lambda )}\ b_{\lambda }~.
\label{eqmotion}
\end{equation}
The commutator in (\ref{eqmotion}) is calculated with the $c$-representation of the Hamiltonian (\ref{fermionH}), and then transformed back to the $b_\lambda$ operators using (\ref{transfunit}). Finally, one collects the corresponding coefficients of both sides of the equation (\ref{eqmotion}). As in a usual eigenvalue problem, we are led to a set of coupled linear equations in finite differences for the wave functions $u_{\lambda n}$, which is an infinite set in the thermodynamics limit. For our problem, we get the set:
\begin{eqnarray}
\left(\epsilon_\lambda^{\mathstrut}-\mu' H\right) u_0^{(\lambda)}+\frac{J'}2 u_1^{(\lambda)}&=&0~, \label{boundary}\\
\left(\epsilon_\lambda^{\mathstrut}-\mu H\right) u_1^{(\lambda)}+\frac{J'}2 u_0^{(\lambda)}+\frac {J}2 u_1^{(\lambda)}&=&0~, \label{nextto}\\
\left(\epsilon_\lambda^{\mathstrut}-\mu H\right) u_n^{(\lambda)}+\frac J2\left(u_{n-1}^{(\lambda)}+u_{n+1}^{(\lambda)}\right)&=&0~,\quad n\geq2~. \label{bulk}
\end{eqnarray}
This set of equations is solved by recurrence methods, yielding the eigenvalues ${\epsilon_\lambda}$ as well as the corresponding wave functions $u_n^{(\lambda)}$~\cite{tsukernik}. General solutions are sought in the form
\begin{equation}
u_{n}=A~r^{n} +B~r^{-n}~,
\label{geral}
\end{equation}
as a linear combination of independent particular solutions $r^n$ and $r^{-n}$. In (\ref{geral}) and subsequent relations, the index $\lambda$ is omitted. Substitution in relation (\ref{bulk}) leads to the spectrum:
\begin{equation}
\epsilon=\mu H-\frac J2\left(r+\frac1{r}\right),
\label{energy}
\end{equation}
which is given in terms of the still unknown quantities $r$ and $r^-1$, related to the wave function (\ref{geral}). The coefficients $A$ and $B$ in (\ref{geral}) can be referred to the initial condition $u_0$ using equations (\ref{boundary}) and (\ref{nextto}).
Two kinds of solutions of (\ref{geral}) are obtained. For complex $r$, one gets a continuous spectrum corresponding to the bulk of the chain, with
\begin{equation}
u_{n}^(\lambda)=A_k~\exp(ikn) +B_k\exp(-ikn)~,
\label{band}
\end{equation}
leading through (\ref{energy}) to the band dispersion:
\begin{equation}
\epsilon(k)=\mu H-J\cos k, \quad 0<k<\pi~.
\end{equation}
Real solutions of (\ref{geral}) are related to bound states, and as such, they should be limited. If we choose $|r|<1$ in (\ref{geral}), we must set $B=0$ and the wave function will decrease exponentially to the interior of the chain. The condition $B=0$ leads to the second degree equation for $r$:
\begin{equation}
\left(1-\eta^2\right)r^2+2\Delta r~+1=0,
\label{2degree}
\end{equation}
with the impurity parameters $\eta$ and $\Delta$ defined in (\ref{parameters}). Depending on values of the impurity parameters, one can split off two, one, or none localized levels from the band. The energy of bound states, when existing, are obtained through formula (\ref{energy}). The different regions of solution in the impurity parameter space are displayed in Fig.~\ref{levels}, with coordinates $\Delta$ and $\eta^2$.

The four characteristic regions of solutions are labeled by the capital letters A, B, C and D. There are no bound states in region A, while there is one bound level in regions B and C, with energy below the band, for $\Delta<0$, and above the band for $\Delta>0$. In region D there are two localized levels, one above and one below the continuous spectrum. Localized states satisfy the condition $|r|<1$, and are real solutions of (\ref{2degree}). Then, it is clear that we obtain at most two localized levels. We now turn to the calculations of the dynamical properties of the chain, when an inhomogeneous state is prepared as initial state. The detailed calculations for all the regions in the parameter space, including behaviors of borders between regions and the crossing at $\Delta=0,~\eta^2=2$, will be published elsewhere \cite{next}. Here, we will focus on a few relevant examples.

\section{Time evolution}
\label{sec:2}
We assume an initial state for the chain in which each `atom' is in one of the two eigenstates of the $S_{n}^{z}$ component of spin $1/2$ (spin up or down). The system initial state then reads
\begin{equation}
\left|\Psi(0)\right\rangle=\left|m_z^{(0)}(0), m_z^{(1)}(0), \dots, m_z^{(N)}(0)\right\rangle,\quad m_z^{(n)}(0)=\pm1/2,\quad n\geq 0.
\label{estin}
\end{equation}
%
%

The above is an eigenstate of $S^z$, the $z$-component of the total spin. Excluding the homogeneous states, (\ref{estin}) is not an eigenstate of Hamiltonian (\ref{ham}), so it displays a non trivial dynamics. To describe the latter, we calculate the \(z\)-axis magnetization for each site \(n\) at an arbitrary instant of time \(t\). Denoting this quantity by \(\left\langle S_{n\vphantom1}^{\mathstrut z}\right\rangle_{\!\!t}\), we have
\begin{equation}
\left\langle S_{n\vphantom1}^{\mathstrut z}\right\rangle_{\!\!t}=\left\langle\Psi(t)\,\middle|\,S_{n\vphantom1}^{\mathstrut z}\,\middle|\,\Psi(t)\right\rangle,
\label{defmagn}
\end{equation}
where \(\left|\Psi(t)\right\rangle=e^{-\frac i\hbar\mathcal H t}\left|\Psi(0)\right\rangle\) is the system state at time \(t\). The Schr\"{o}dinger expectation value $\left\langle S_{n\vphantom1}^{\mathstrut z}\right\rangle$ can be written in the Heisenberg picture,
where the average is taken over the initial state.
The Heisenberg operator \(S_{n\vphantom1}^{\mathstrut z}(t)\) can be expressed in terms of the quasiparticle operators \(b_\lambda^{\mathstrut}\) and \(b_\lambda^\dag\), whose trivial time evolution
\begin{equation}
b_\lambda^{\mathstrut}(t)=b_\lambda^{\mathstrut} e^{-\frac i\hbar\epsilon_\lambda^{\mathstrut} t},\quad b_\lambda^\dag(t)=b_\lambda^\dag e^{\frac i\hbar\epsilon_\lambda^{\mathstrut} t},
\label{bsHeins}
\end{equation}
allows us to obtain the explicit time dependence of the magnetization. Subsequently, using the result for arbitrary \(t\), we address the question of obtaining the magnetization for very large values of time, i.e., long-time tails. In the thermodynamic limit ($N\rightarrow\infty$), this is done using the stationary phase method~\cite{SPM}, to estimate the dominant behavior at asymptotically long times. Once the initial state is given in the form (\ref{estin}), with $m_z^{(n)}(0)$ being the $n$-spin component of the initial state, the site magnetization can be written as:
\begin{equation}
\left\langle S_{n\vphantom1}^{\mathstrut z}\right\rangle_{\!\!t}=\sum_{m=0}^Nm_z^{(m)}(0)S_{n,m}(t),\quad n\geq 0~,
\label{3rdresger}
\end{equation}
where the interference effects are contained in the factor $S_{n,m}(t)$, given below
\begin{equation}
S_{n,m}(t)=\left|\sum_{\lambda=1}~u_{n\vphantom1}^{(\lambda)} {u_{m\vphantom1}^{(\lambda)}}^*e^{-\frac i\hbar\epsilon_\lambda^{\mathstrut} t}\right|^2~,
\label{Snm}
\end{equation}
with \{$u_{m}^{\lambda}$\} being the wave functions obtained through the diagonalization process (see Eq. (\ref{transfunit})). In the general case, the sum (\ref{Snm}) includes the continuous as well as the discrete spectra. Exact asymptotic series can be obtained for the long-time behavior of $S_{n,m}(t)$. In general, dominant terms of the series come from contributions of stationary points and they show a slow relaxation, in the form of a power law $t^{-\alpha}$, with $\alpha\geq1$~\cite{next}. There are remarkable exceptions that we want to discuss in the next Section.
\section{Results and discussion}
\label{sec:3}
Without loss of generality, we will choose a particular case of (\ref{estin}) as initial state:
\begin{equation}
\left|\Psi(0)\right\rangle=\left|\uparrow\right\rangle_{\!0}\left|\downarrow\right\rangle_{\!1}\left|\uparrow\right\rangle_{\!2}\left|\uparrow\right\rangle_{\!3}\dots\left|\uparrow\right\rangle_{\!N}~.
\label{specific}
\end{equation}
The only spin reversed in the initial configuration is at site $n=1$, the spin next to the impurity. In spite of this simple structure, the initial state has a moderate energy distribution, being a linear combination of many eigenstates of the system, a prerequisite for typicality \cite{reimann,tasaki}. Since the state is biased near the impurity, it will have non vanishing admixtures of the localized levels, when they are present in the spectrum. In our approach, the initial state (\ref{specific}) is kept fixed, but we change the parameters of the model, thus changing its energy distribution.
We present results corresponding to the time evolution for values of parameters in the different regions of Fig.~\ref{levels}.
\subsection{Region A, no impurity levels}
\label{ssec:1}
\begin{figure}[ht!]
\includegraphics [width=\linewidth]{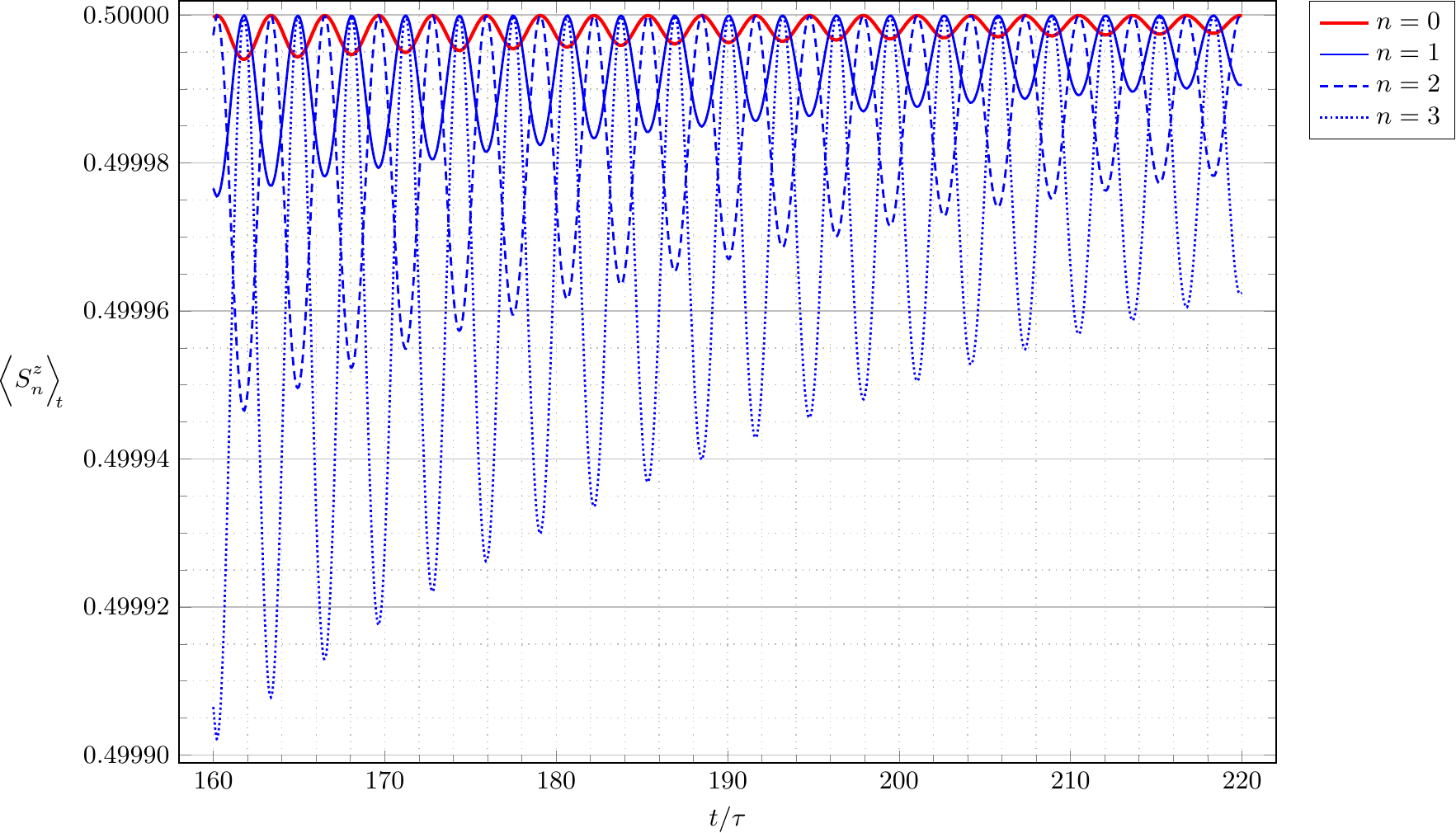}
\caption{Long-time behavior of the magnetization at sites $n=0,1,2,3$, for parameters in region A of Fig.~\ref{levels}. The example shown corresponds to parameters \(\Delta=0\), \(\eta^2=1\), which is the pure system, but the behavior is typical for all values of parameters within region A. The unit of time is given by \(\tau=\hbar/\left|J\right|\).}
\label{plotregA}
\end{figure}

For region A, there are no bound states. The relaxation is driven by states within the continuum, where the interference is in general, destructive. There are two stationary points, which correspond to the edges of the band, which yield a slower evolution in the form of a damped oscillation that relaxes as $ 1/t^3$. This slow relaxation is dominant asymptotically, and is
due to constructive interference of states in the neighborhood of stationary points. The frequency of the oscillation is the band width, $\omega=2J/\hbar$. In Fig.~\ref{plotregA}, we display the long-time evolution, for the first sites of the chain ($n=0,1,2,3$).

As displayed in the figure, at asymptotically long-times, all oscillations have the same frequency and are damped in time. For the time scale shown in the figure, the amplitude of the oscillations is already very small. This slow relaxation follows a power law ($ 1/t^3$), which is characteristic of quantum behavior. The above time dependence has been found before in \cite{tjon}. The magnetization of the extrapolated final state is homogeneous and stationary. This is the picture we intuitively expect for the \emph{equilibrium state} in typical systems.
\subsection{Regions B or C, one impurity level}
\label{ssec:2}
\begin{figure}[ht!]
\includegraphics[width=\linewidth]{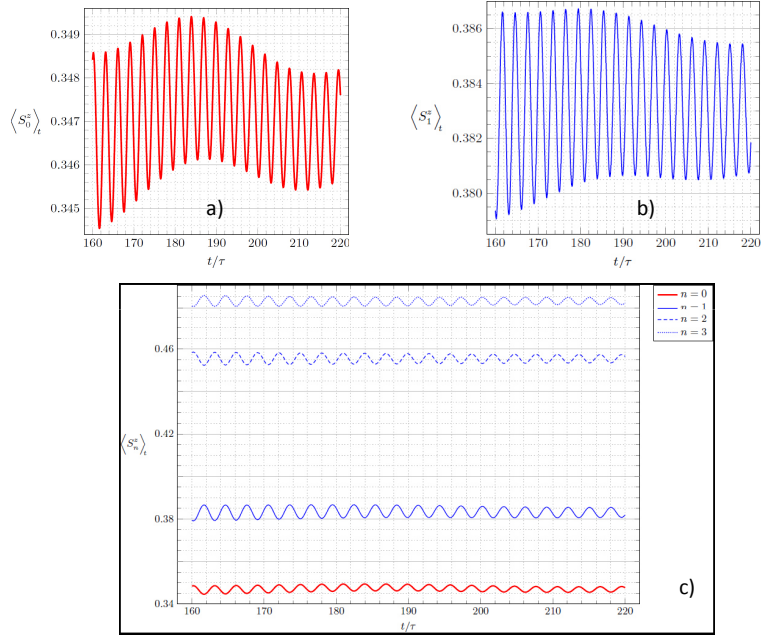}
\caption{Long-time behavior of the magnetization for regions B or C. The two frequencies of the asymptotic oscillations are apparent in the upper panels of the figure, a) for site $n=0$, and b) for $n=1$. One notes that a fast oscillation is accompanied by a low frequency modulation. Values of parameters are \(\left|\Delta\right|=1/2\), \(\eta^2=2\). In the lower panel c), we show in a different vertical scale, the magnetization for the sites $n=0,1,2,3$, to visualize the asymptotic limits. Relaxation is very slow, and only the oscillation with the `high frequency' is evident. The unit of time is given by \(\tau=\hbar/\left|J\right|\).}
\label{regionBC}
\end{figure}

In regions B or C, one bound state is split off from the band. There is a symmetry between the two subregions with $\Delta>0$ and $\Delta<0$, where the localized state splits from above or below the band, respectively (this assumes $J>0$; if $J$ changes sign, the statement is reversed).
For the dynamics, the sign of $J$ makes no difference, and the asymptotic oscillatory term has two different frequencies superposed. Now the time evolution of the magnetization shows a peculiar property, with oscillations decaying as $1/t^{3/2}$ to a \emph{constant value} which depends on the site. The two frequencies that appear in the long-time behavior are given by the difference between the localized level and the edges of the band:
\begin{equation}
\omega_{\pm}=\frac{(\epsilon_i-\mu H) \pm J}{\hbar}~,
\label{frequencies}
\end{equation}
where $\epsilon_i$ is the energy of the bound state. Thus, one gets a `low' and a `high' frequency oscillation in the asymptotic behavior. An example is shown in Fig.~\ref{regionBC}, for the parameter values $|\Delta|=1/2$ and $\eta^2=2$. There is only one acceptable root of (\ref{2degree}), with value $r=(1-\sqrt{5})/2$, corresponding to the energy $\epsilon=\mu H \pm J\sqrt{5}/2$ of the bound state, depending on the sign of $\Delta$, \emph{i.e.} a localized level above the band for $\Delta=1/2$, and below the band for $\Delta=-1/2$. Frequencies of the asymptotic oscillations are $(J/2)(\sqrt{5}\pm2)/\hbar$, and both oscillations are affected by a factor that decays in time as $t^{-3/2}$. This comes from the exact result for the asymptotic series.

In the thermodynamic limit, one can think that any part of the interacting system is coupled to an infinite reservoir consisted of the complementary part. Intuitively, one expects that any initial inhomogeneity will be wiped out by the interactions, and the time evolution will drive the system to a final equilibrium state, that will be stationary and homogeneous. In the present case, the magnetization extrapolated at $t\rightarrow \infty$ is stationary, but inhomogeneous, showing a site dependent profile near the impurity. We consider this unusual behavior as non-ergodic.

\subsection{Region D, two impurity levels}
\label{ssec:3}
\begin{figure}[ht!]
\includegraphics[width=\linewidth]{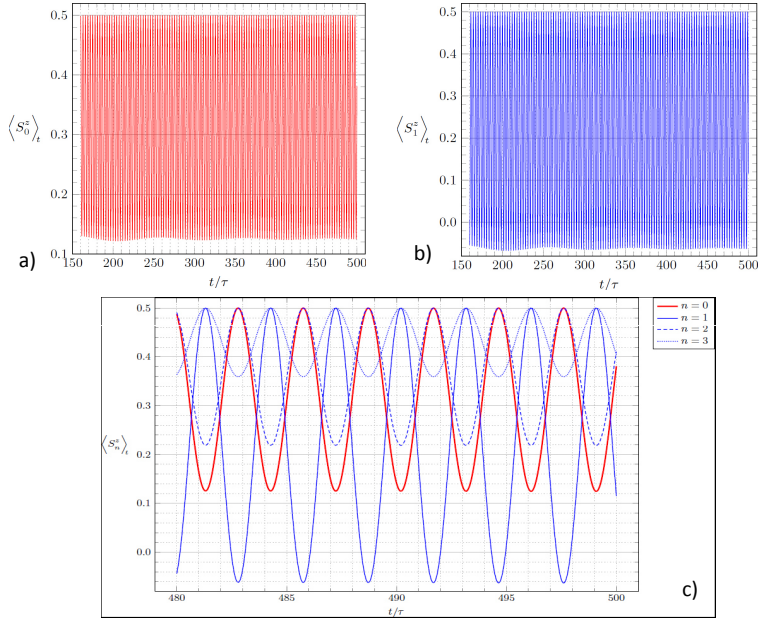}
\caption{Long-time behavior of the magnetization for region D, where two localized levels are split off from the band. Values of parameters are \(\Delta=0\), \(\eta^2=3\). The asymptotic oscillations involved three frequencies, as shown in the upper panel: a) for site $n=0$; b) for site $n=1$. In the lower panel c), we show in a different time scale, the magnetization for the sites $n=0,1,2,3$, to visualize the undamped oscillations. Due to the scale, only the Rabi frequency is apparent. The unit of time is given by \(\tau=\hbar/\left|J\right|\).}
\label{regionD}
\end{figure}

Region D is the characteristic region of parameter space with the presence of two bound states, one above and one below the continuous band. The stationary phase methods yields the dominant terms of the asymptotic series in the form of oscillations with five different frequencies, two `low', two `high', and the highest which corresponds to the Rabi frequency of the two bound states, $\omega_{Rabi}=|\epsilon_1-\epsilon_2|/\hbar$~. The `low' and `high' frequency oscillations are similar to those obtained in subsection \ref{ssec:2} for the case of one impurity state, except that here we have two localized levels. When $\Delta=0$, the levels split symmetrically from the band edges, and the number of frequencies is reduced to three. The `high' and `low' frequency oscillations are damped in time by a factor proportional to $1/t^{3/2}$, as in the previous case. An example is shown in Fig.~\ref{regionD}, for values of parameters $\Delta=0$ and $\eta^2=3$. In Fig.~\ref{regionD}a) and Fig.~\ref{regionD}b), we display the asymptotic oscillations for sites $n=0$ and $n=1$, respectively. The figures are rather intricate, since they are obtained as a superposition of oscillations with three different frequencies: the Rabi frequency $(3/\sqrt{2})(J/\hbar)$, the `high' frequency $(1+3/\sqrt{8})(J/\hbar)$, and the `low' frequency $(1-3/\sqrt{8})(J/\hbar)$. The highest frequency (Rabi) oscillation is resolved in Fig.~\ref{regionD}c), with a different time scale.

What is \emph{remarkable} in Figure (\ref{regionD}c) is the fact that the Rabi oscillation is not damped in time. Note that the amplitude of the undamped oscillation is huge at the first sites of the chain; the magnetization for $n=1$ even takes negative values. Oscillations with frequencies different from the Rabi frequency are still present, but their amplitude decays in time as $1/t^{3/2}$, as stated above, and asymptotically their contribution can be ignored. Therefore, when two impurity levels exist, the quantum interference between them leads to oscillations which settle at very long times with no damping, since localized states are unable to `thermalize' with states of the continuum spectrum. The extrapolated magnetization at $t~\rightarrow~\infty$ is never stationary, nor homogeneous.
\subsection{Final discussion}
\label{ssec:4}
In summary, the time evolution of a typical initial state is driven by quantum fluctuations and interference effects among the many energy eigenstates that participate in its linear combination. In the thermodynamic limit quantum recurrence times are very long (infinite), and one can argue that parts of the system act as `reservoirs' for other parts. Under these conditions, one expects the system to relax to the equilibrium state. An underlying assumption in the above discussion, is the presence of a continuous energy spectrum in the thermodynamic limit. This behavior is observed in our model, when impurity parameters are within region A, in Figure \ref{levels}. Relaxation leads to a state where the magnetization is stationary and homogeneous, which we identify as the equilibrium state. Non-ergodicity appears when we adequately change the impurity parameters over the different regions of solution. This process changes the energy spectrum, the admixtures of the energy eigenstates, and the energy distribution of the initial state. Admixture of localized states leads to non-ergodic behavior, as shown in the previous Subsections. When only one level is split off from the continuum (regions B and C), the system relaxes to a state with nonhomogeneous magnetization. In the presence of two localized levels (region D), the system never relaxes. The quantum interference between the two bound states yields undamped magnetization oscillations with the Rabi frequency of the levels. This relation between non-ergodicity and localization has been noted before. In Ref. \cite{localization}, the effect has been discussed for the localization-delocalization transition. This is an extreme case where all energy levels are localized in the non-ergodic phase. Our example shows that splitting a few localized levels from the continuum is enough to get non-ergodic behavior.

\section*{Acknowledgements} \label{Sec-acknow}

The authors are grateful to S\~ao Paulo Research Foundation (\textbf{FAPESP}, Brazil) for financial support through project No. 2009/53826-8. GGC also acknowledges support from project \textbf{FAPESP} 2017/07016-0.

\section*{Author contribution statement}
The two authors contributed equally in the elaboration of the present paper.

%

%
%

\end{document}